\documentclass[aps,pra,reprint,superscriptaddress,longbibliography]{revtex4-1}
\usepackage{graphicx,color}
\usepackage{amsmath, amssymb}
\usepackage{longtable}
\usepackage{natbib}
\usepackage{bm}
\usepackage{hyperref}

\allowdisplaybreaks[1]

\begin{document}
\title{Manifestation of vibronic dynamics in infrared spectra of Mott insulating fullerides}
\author{Yuki Matsuda}
\affiliation{Theory of Nanomaterials Group, University of Leuven, Celestijnenlaan 200F, B-3001 Leuven, Belgium}
\affiliation{Department of Physics, Graduate School of Science, Tohoku University, 6-3 Aramaki, aza-Aoba, Aoba-ku, Sendai 980-8578, Japan}
\author{Naoya Iwahara}
\email{naoya.iwahara@gmail.com}
\affiliation{Theory of Nanomaterials Group, University of Leuven, Celestijnenlaan 200F, B-3001 Leuven, Belgium}
\author{Katsumi Tanigaki}
\affiliation{Department of Physics, Graduate School of Science, Tohoku University, 6-3 Aramaki, aza-Aoba, Aoba-ku, Sendai 980-8578, Japan}
\affiliation{WPI-Advanced Institute for Materials Research (WPI-AIMR), Tohoku University, 2-1-1 Katahira, Aoba-ku, Sendai, 980-8577, Japan}
\author{Liviu F. Chibotaru}
\email{liviu.chibotaru@gmail.com}
\affiliation{Theory of Nanomaterials Group, University of Leuven, Celestijnenlaan 200F, B-3001 Leuven, Belgium}
\date{\today}

\begin{abstract}
The fine structure and temperature evolution of infrared spectra have been intensively used to probe the nature of Jahn-Teller dynamics in correlated materials. 
At the same time, theoretical framework to adequately extract the information on the complicated vibronic dynamics from infrared spectra is still lacking. 
In this work, the first-principles theory of the infrared spectra of dynamical Jahn-Teller system is developed and applied to the Mott-insulating Cs$_3$C$_{60}$.
With the calculated coupling parameters for Jahn-Teller and infrared active vibrational modes, the manifestation of the dynamical Jahn-Teller effect in infrared spectra is elucidated. 
In particular, the temperature evolution of the infrared line shape is explained. 
The transformation of the latter into Fano resonance type in metallic fulleride is discussed on the basis of obtained results. 
\end{abstract}

\maketitle

\section{Introduction}
The role of dynamical Jahn-Teller (JT) effect \cite{Bersuker1989, Kaplan1995, Chancey1997} in electronic property of correlated materials attracts significant attention.
Such materials of current interest include
e.g. various transition metal compounds \cite{Krimmel2005, deVries2010, Nakatsuji2012, Kamazawa2017, Nirmala2017}, rare-earth \cite{Webster2007} and actinide dioxides \cite{Santini2009}. 
An increasing precision of the investigation of such materials demands detailed knowledge of the nature of JT dynamics and the mechanisms of its manifestation in observed properties. 
Rich fine structures in infrared (IR) spectra are expected to encode the information on the local JT dynamics. 
Despite the fact that IR spectroscopy has been applied to various systems \cite{Yamaguchi1997, Klupp2006, Jung2008, Francis2012, Klupp2012, Kamaras2013, Qu2013, Zadik2015, Constable2017, Lavrentiev2017}, the relation between the JT dynamics and the structure of IR spectra has not been established.
One family of materials where the JT effect has been much investigated are the alkali-doped fullerides \cite{Gunnarsson1997, Gunnarsson2004, Auerbach1994, Manini1994, OBrien1996, Chibotaru2005, Klupp2012, Kamaras2013, Zadik2015, Nomura2016, Nava2018, Iwahara2018}.

Recently, the understanding of the dynamical JT effect on the C$_{60}^{3-}$ sites of alkali-doped fullerides $A_3$C$_{60}$ ($A =$ K, Rb, Cs) has significantly advanced.
The accurate calculation of the orbital vibronic coupling parameters for C$_{60}$ anions \cite{Iwahara2010} enables to access realistic low-energy vibronic spectra of C$_{60}^{n-}$ ($n = $ 1-5) \cite{Iwahara2013, Liu2018}. 
Since the dynamical JT (vibronic) state can be thought of as a quantum superposition of statically JT deformed molecular wave functions \cite{Judd, Cibotaru},
it is simultaneously characterized by the presence of the JT split adiabatic orbitals and the equal contributions of the degenerate electronic states to it. 
Therefore, the presence of the unquenched JT dynamics in Mott-insulating phase due to a large dynamical JT 
stabilization energy naturally explains the absence of orbital ordering and the isotropic character of the antiferromagnetic exchange interaction
\cite{Chibotaru2005, Iwahara2013}. 
The dynamical JT effect was found not to be quenched by band effects in the metallic phase \cite{Iwahara2015}, a fact confirmed indirectly by experiment \cite{Zadik2015}.
Since each C$_{60}^{3-}$ site has nondegenerate three adiabatic orbitals, the conduction band of $A_3$C$_{60}$ is also split into three inequivalent subbands. 
As a consequence, the electron correlation basically develops only in one JT split half-filled adiabatic subband
(see Fig. 2d in Ref. \cite{Iwahara2016}),
and hence the critical intrafullerene electron repulsion for the JT-induced orbital-selective Mott transition is small
\cite{Iwahara2015}
\footnote{
In the inequivalent subbands, the electron correlation develops in different ways. 
The evolution of the electron correlation can be seen in the Gutzwiller's reduction factors with respect to Coulomb repulsion on C$_{60}$ sites (see Fig. 2d in Ref. \cite{Iwahara2016}).
As the increase of the Coulomb repulsion on C$_{60}$ sites the Gutzwiller's factor of half-filled subband becomes close to zero (strongly correlated), whereas those for empty and filled subbands become close to one (uncorrelated). 
Since the metallic screening and the band energy for nondegenerate system are smaller than those of degenerate band, the critical Coulomb repulsion on C$_{60}$ sites for the Mott transition is smaller too (Sec. VI in Ref. \cite{Iwahara2015}).
},
which is consistent with the recent estimate of Coulomb repulsion by electrical conductivity measurements \cite{Matsuda2018}.
Based on the derived vibronic structure of fullerene anions, it was also shown that the spin-gap from NMR data \cite{Brouet2002c, Jeglic2009} 
is
well reproducible \cite{Liu2018}.

The IR measurements of $A_3$C$_{60}$ have been performed across the Mott transition.
In the Mott insulating phase, the temperature dependence of the IR spectra was attributed to the evolution of the JT dynamics due to the variation of the structure of the adiabatic potential energy surface 
by thermal expansion of the crystal \cite{Klupp2012, Kamaras2013}.
The spectral shape in metallic phase close to the Mott transition remains similar to that  
in Mott insulating phase,
whereas it gradually changes to Fano resonance type as departing from it \cite{Zadik2015}. 
The unchanged shape of the IR spectrum in the vicinity of Mott transition was interpreted as the evidence of the JT effect on C$_{60}^{3-}$, following the theoretical prediction \cite{Iwahara2015}.
On the other hand, the evolution of the Lorenzian shape into Fano resonance deep in metallic phase was interpreted in Ref. \cite{Zadik2015} as a quenching of JT effect, which contradicts the theoretical 
calculations \cite{Iwahara2015}
\footnote{
The calculations in Ref. \cite{Iwahara2015} show unquenched dynamical JT effect on the C$_{60}^{3-}$ sites in Cs$_3$C$_{60}$ which is gradually quenched approaching the limit of JT glass in fullerides of a smaller volume like K$_3$C$_{60}$. 
The observed transformation of the shape of the IR line can be related to this change of the character of JT effect but not to its full disappearance as was supposed in Ref. \cite{Zadik2015}.
See also Sec. \ref{Sec:Discussion}.
}
and NMR \cite{Brouet2002c, Alloul2017} 
and electrical conductivity \cite{Matsuda2018} measurements: 
The NMR measurements show the spin gap induced by the JT dynamics in metallic Rb$_3$C$_{60}$ \cite{Brouet2002c} and 
the same magnetic properties in both Mott insulating phase and high temperature paramagnetic insulating phase above metallic one \cite{Alloul2017}
and the electrical conductivity measurements show no drastic change of resistivity in the whole metallic domain \cite{Matsuda2018}.
In order to correctly assess the nature of JT dynamics from the fine structure of IR spectra, it is decisive to develop a theoretical framework for their adequate treatment. 

In this work, a fully quantum mechanical theory for IR spectra of dynamical JT systems is developed. 
Combining the developed theory and first-principles calculations, the IR active vibronic states of Cs$_3$C$_{60}$ cluster are derived and, on this basis, IR spectra are simulated. 
In terms of the obtained vibronic states, the relation between the JT dynamics and the temperature evolution of IR spectra is revealed. 
The developed theoretical scheme will be indispensable for the understanding of the phenomena related to the photoexcitation processes in fullerides and other dynamical JT materials.

\section{Model vibronic Hamiltonian of C$_{60}^{3-}$} 
The minimal model describing the manifestation of the JT dynamics in IR spectra is derived. 
Below, the irreducible representations (irrep) of $I_h$ group are used, the components of the irreps. are always real in the main text, and the lower (upper) case is used for the irrep of the one electron orbital state and the mass-weighted normal mode (electronic term and vibronic states).
Three orthogonal $C_2$ axes are chosen as the Cartesian axes (see Fig. S1 \cite{SM}). 

In the low-energy states of C$_{60}^{3-}$ site in $A_3$C$_{60}$, triply degenerate $t_{1u}$ lowest unoccupied molecular orbitals (LUMO) are half-filled because of the deep LUMO levels and large separation from the other levels.
The bielectronic interaction between electrons occupying the LUMOs, $\hat{H}_\text{bi}$, induces the term splitting as ${}^4A_u \oplus {}^2T_{1u} \oplus {}^2H_u$.
$\hat{H}_\text{bi}$ raises the $^2T_{1u}$ term by $2J_\text{H}$ with respect to the $^2H_u$ term, where $J_\text{H}$ is the Hund's rule coupling parameter ($\approx 40$ meV \cite{Nomura2012, Iwahara2013, Liu2018}). 
The $t_{1u}$ orbitals couple to the totally symmetric $a_g$ and the five-fold degenerate $h_g$ vibrations of the C$_{60}$ cage, and the linear vibronic coupling to the $h_g$ modes, $\hat{H}^{(1)}_\text{vibro}$, gives rise to the JT effect \cite{Auerbach1994, OBrien1996, Bersuker1989, Chancey1997}. 

Within the simplest model for C$_{60}^{3-}$ consisting of $\hat{H}_\text{bi}$, the Hamiltonian of harmonic oscillations, $\hat{H}_0$, and $\hat{H}_\text{vibro}^{(1)}$, 
\begin{eqnarray}
 \hat{H}^{(0)} &=& \hat{H}_\text{bi} + \hat{H}_0 + \hat{H}_\text{vibro}^{(1)},
\label{Eq:H0}
\end{eqnarray}
the IR active $t_{1u}$ vibrational degrees of freedom are uncoupled from the JT dynamics. 
Therefore, neither fine structure nor temperature evolution in IR spectra can be expected.
In order to reveal the role of the JT dynamics in the IR spectra, the interplay of the vibronic and the IR vibrational degrees of freedom has to be considered. 
The interplay arises from nonlinear vibronic couplings admixing the IR modes. 
Although the $t_{1u}$ LUMOs do not linearly couple to the IR active $t_{1u}$ modes, they do to the products of the coordinates when the latter include the $h_g$ representation (see Table S2 \cite{SM}).
Thus, 
the minimal model should contain the quadratic vibronic term to the IR active modes allowing to indirectly relate the IR and JT modes via electronic orbitals,
\begin{eqnarray}
 \hat{H} &=& \hat{H}^{(0)} + \hat{H}^{(2)}_\text{vibro}, 
\label{Eq:H}
\end{eqnarray}
where, 
\begin{eqnarray}
 \hat{H}^{(2)}_\text{vibro} &=& 
  \sum_{\mu_i \Gamma_i} \sum_{\nu \gamma} 
  \sum_{\lambda \lambda'} 
 V^{\Gamma_1 \Gamma_2}_{\nu \mu_1 \mu_2}
 \{\hat{q}_{\Gamma_1(\mu_1)} \otimes \hat{q}_{\Gamma_2(\mu_2)} \}_{\nu h_g\gamma}
\nonumber\\
 &\times& 
 \sqrt{\frac{15}{2}}
 \langle t_{1u} \lambda|h_u \lambda', h_g \gamma \rangle
  |T_{1u} \lambda \rangle 
  \langle H_u \lambda'|
 + h.c.
\label{Eq:HQJT}
\end{eqnarray}
Here, $|T_{1u}\lambda\rangle$ ($\lambda = x,y,z$) and $|H_u\lambda'\rangle$ ($\lambda' = \theta, \epsilon, \xi, \eta, \zeta$ 
which stand for $(2z^2-x^2-y^2)/\sqrt{6}$, $(x^2-y^2)/\sqrt{2}$, $\sqrt{2}yz$, $\sqrt{2}zx$, $\sqrt{2}xy$, respectively) 
are electronic term states, 
$\hat{q}_{\Gamma \gamma}$ is nuclear mass-weighted normal coordinate operator
\footnote{Displacement $u_{A\lambda}$ of nuclei $A$ along $\lambda$ ($= x,y,z$) direction from a high-symmetric reference structure is expressed by mass-weighted coordinates as \unexpanded{$u_{A\lambda} = \sum_{\mu \Gamma \gamma} q_{\Gamma(\mu)\gamma} e_{A\lambda}^{\Gamma(\mu)\gamma}/\sqrt{M_A}$}, where $M_A$ is the mass of nucleus $A$, and \unexpanded{$e_{A\lambda}^{\Gamma(\mu)\gamma}$} is the polarization vector of $\Gamma(\mu)\gamma$ normal mode.
The polarization vector is dimensionless, and the dimension of $q_{\Gamma \gamma}$ is $\sqrt{m_e}a_0$, where $m_e$ is mass of electron and $a_0$ is Bohr radius. 
See for detail Sec. 10.1 in Ref. \cite{Inui1990}.
},
$V^{\Gamma_1 \Gamma_2}_{\nu \mu_1 \mu_2}$ is the quadratic orbital vibronic coupling parameter,
$\{\hat{q}_{\Gamma_1(\mu_1)} \otimes \hat{q}_{\Gamma_2(\mu_2)} \}_{\nu h_g\gamma}$ is the symmetrized product of corresponding nuclear 
coordinates [see Eq. (\ref{Eq:qq}) and Ref.\cite{Bersuker1989}],
$\nu$ distinguishes multiple $h_g$ representations since $I_h$ is not simply reducible group, and 
$\langle t_{1u} \lambda|h_u \lambda', h_g \gamma \rangle$ is the Clebsch-Gordan coefficient (Sec. I in Ref. \cite{SM}). 
The coefficient $\sqrt{15/2}$ in Eq. (\ref{Eq:HQJT}) is introduced as in 
Refs. \cite{Auerbach1994, OBrien1996, Chancey1997}.
The absence of the diagonal block in $\hat{H}_\text{vibro}^{(2)}$ is due to the seniority selection rule \cite{Racah1943}.
In $\hat{H}_\text{vibro}^{(2)}$, the quadratic JT coupling is not included because it does not influence much the vibronic states of C$_{60}^{3-}$ \cite{Liu2018Q}.
See for the derivation of Eq. (\ref{Eq:HQJT}) Appendix \ref{A:derivation}.

\begin{table}[tb]
\caption{
Calculated quadratic vibronic coupling parameters $V^{\Gamma_1\Gamma_2}_{\nu \mu_1 \mu_2}$ ($\times 10^{-7}$ a.u.).
}
\label{Table:W}
\begin{ruledtabular}
\begin{tabular}{ccccccc}
$\Gamma_1(\mu_1)$ & $\Gamma_2(\mu_2)$ & $V^{\Gamma_1\Gamma_2}_{\mu_1 \mu_2}$ &
$\Gamma_1(\mu_1)$ & $\Gamma_2(\mu_2)$ & $\nu$ & $V^{\Gamma_1\Gamma_2}_{\nu \mu_1 \mu_2}$ \\ 
\hline
$t_{1u}(4)$ & $t_{1u}(4)$ & $-3.57$  & $g_u(5)$   & $h_u(6)$    & 1 &   1.19  \\
            & $g_u(5)$    &   3.19   &            &             & 2 &   1.60  \\
            & $g_u(6)$    &   0.38   & $g_u(6)$   & $g_u(6)$    & - &   1.46  \\
            & $h_u(6)$    &   6.18   & $h_u(6)$   & $h_u(6)$    & 1 &   2.16  \\
$g_u(5)$    & $g_u(5)$    & $-1.89$  &            &             & 2 &   3.23  \\
\end{tabular}
\end{ruledtabular}
\end{table}

\section{Nonlinear vibronic coupling constants}
The frequencies $\omega_{\Gamma(\mu)}$ and orbital vibronic coupling parameters $V^{\Gamma_1 \Gamma_2}_{\nu \mu_1 \mu_2}$ 
of Cs$_3$C$_{60}$ clusters were calculated using density functional theory (DFT) with hybrid exchange-correlation functional (Appendix \ref{A:DFT}).
Since the IR peak of the highest frequency mode ($\omega_{t_{1u}(4)} \approx$ 1360 cm$^{-1}$) has the richest fine structure \cite{Klupp2012, Kamaras2013, Zadik2015}, we will describe only this peak. 
The obtained parameters for A15 and fcc lattices of Cs$_3$C$_{60}$ are similar to each other, hence
the parameters for A15 are used below. 
The calculated frequencies for neutral C$_{60}$ are in line with experimental data (Table S3 \cite{SM}).
Upon doping, the frequency of the $t_{1u}(4)$ mode shows relatively large red shift by about 60 cm$^{-1}$ and approaches to the frequencies of $g_u(5)$ and $h_u(6)$ modes, $\omega_{t_{1u}(4)} = 1384$, $\omega_{g_u(5)} = 1334$, and $\omega_{h_u(6)} = 1351$ cm$^{-1}$. 
On the other hand, the $g_u(6)$ mode whose frequency is close to that of $t_{1u}(4)$ in the neutral C$_{60}$, implying a relevance to the spectra \cite{Klupp2012}, does not vary much, $\omega_{g_u(6)} = 1440$ cm$^{-1}$.
This result points to the importance of $t_{1u}(4)$, $g_u(5)$, and $h_u(6)$ rather than $g_u(6)$ for the description of the IR spectrum because these IR and JT inactive modes ($g_u$, $h_u$) may couple nonlinearly to the $t_{1u}$ orbitals (see for the polarization vectors Fig. S1 \cite{SM}).

The quadratic orbital vibronic coupling parameters $V^{\Gamma_1 \Gamma_2}_{\nu \mu_1 \mu_2}$ (a.u.)
were derived by fitting the $t_{1u}$ LUMO levels with respect to the $t_{1u}z, g_uz, h_u\theta$ deformations (a.u.) 
to the model vibronic Hamiltonian 
(see for the model Hamiltonian Appendix \ref{A:HQJT}, for the fitting Fig. S2 in Ref. \cite{SM}, and for the coupling parameters Table \ref{Table:W}). 
One should note that the magnitudes of $V^{\Gamma_1 \Gamma_2}_{\nu \mu_1 \mu_2}$, not only for the IR active $t_{1u}(4)$ mode but also for the IR inactive $g_u(5)$ and $h_u(6)$ modes, are similar to each other
\footnote{
The orders of the obtained quadratic coupling parameters $V$ and the expectation value of the product of IR coordinates $\hat{q}^2$ are $10^{-7}$ a.u. and $10^2$ a.u., respectively, therefore, the energy scale of each quadratic vibronic coupling \unexpanded{$\sqrt{15/2} V \langle \hat{q}^2 \rangle$} is $10^{-5}$-$10^{-4}$ a.u. 
Here, \unexpanded{$\langle \hat{q}^2 \rangle$} is approximated as \unexpanded{$(\hslash/\omega) (\langle \hat{n} \rangle + 1/2)$} using the harmonic oscillator model with $\omega \approx 1350$ cm$^{-1}$ and the averaged vibrational quantum number \unexpanded{$\langle \hat{n} \rangle =$ 0-1}.
On the other hand, the energy scale of the linear vibronic coupling is $10^{-3}$ a.u. in terms of the JT energy for C$_{60}^-$.
}.
Indeed, the largest parameter is obtained for the $t_{1u}$ and $h_u$ modes. 
On the other hand, the intermode coupling involving the $t_{1u}(4)$ and $g_u(6)$ is found to be weak.
Given the relatively large difference in frequency between them, the contribution from the $g_u(6)$ mode to the vibronic states is negligible. 
Thus, hereafter, we consider only the $t_{1u}(4)$, $g_u(5)$ and $h_u(6)$ modes which we call IR modes for simplicity 
\footnote{Both the IR active $t_{1u}(4)$ and inactive $g_u(5)$ and $h_u(6)$ modes are called IR modes because these three modes are admixed by the quadratic vibronic coupling, resulting in the contribution to the fine structure of the IR spectrum.}
as well as the JT active $h_g$ modes.

\section{Vibronic states}
\label{Sec:vibronic}
Comparing the energy scales of the linear and quadratic vibronic coupling, the latter is regarded as perturbation to the former \cite{Note4}.
The eigenstates of the unperturbed Hamiltonian, Eq. (\ref{Eq:H0}),
are the direct products of the linear JT state
\footnote{
In this work, we treat two types of vibronic states. 
The first ones are the eigenstates of linear $t_{1u}^3 \otimes 8h_g$ JT Hamiltonian and the second are eigenstates of the full Hamiltonian $\hat{H}$, including nonlinear vibronic couplings to the IR modes. 
To avoid the confusion, we call the former JT states and the latter vibronic states. 
},
$|\Phi^{(0)}_{\kappa}\rangle$, and of the harmonic oscillations of the IR modes \cite{Note5},
$|\bm{n}'\rangle$.
The JT states $\kappa$ are characterized by the irrep $\Gamma$ (or vibronic angular momentum $J$), its component $\gamma$, parity $P$ originating from the seniority, and principal quantum number $\alpha$ distinguishing the energy levels, $\kappa = (\alpha \Gamma \gamma P)$ \cite{Iwahara2013, Liu2018}.
The ground $(\Gamma = T_{1u}, P = +1)$ and the first excited $(H_u, -1)$ JT states are separated by about 8 meV and the other JT levels appear at $>$ 30 meV.

The JT dynamics modifies the strengths of the nonlinear vibronic terms,
\begin{eqnarray}
 \hat{H}_\text{vibro}^{(2)}
 &=&
 \sum_{\mu_i \Gamma_i}
 \sum_{\nu \gamma}
 \sum_{\kappa \kappa'}
 V^{\Gamma_1 \Gamma_2}_{\nu \mu_1 \mu_2}
 \Xi(\kappa; \kappa')
 \{\hat{q}_{\Gamma_1(\mu_1)} \otimes \hat{q}_{\Gamma_2(\mu_2)} \}_{\nu h_g\gamma}
\nonumber\\
&\times&
 \sqrt{\frac{15}{2}} \langle \Gamma_\kappa \gamma_\kappa|\Gamma_{\kappa'} \gamma_{\kappa'}, h_g \gamma \rangle
 |\Phi^{(0)}_{\kappa}\rangle \langle \Phi^{(0)}_{\kappa'}|,
\label{Eq:Hquad}
\end{eqnarray}
where, the electronic basis in Eq. (\ref{Eq:HQJT}) is replaced by the JT states (see Appendix \ref{A:MatHQJT}).
The form remains the same as Eq. (\ref{Eq:HQJT}) except for the factor $\Xi(\kappa;\kappa')$ which is a vibronic factor modifying the operators (see e.g., Refs. \cite{Bersuker1989, Kaplan1995, Chancey1997}).
For the lowest two JT states, $\Xi = 0.63$ \cite{Liu2018}, and thus, the quadratic vibronic coupling is reduced (for other $\Xi$'s, see Table S4 \cite{SM}). 
Besides the reduction factor, the parity selection rule reduces the effect of the quadratic coupling because the coupling is only operative between the linear JT
states with different parities (see Sec. IV B 3 in Ref. \cite{Liu2018}). 

The eigenstates of the full Hamiltonian, Eq. (\ref{Eq:H}), which we call vibronic states \cite{Note6}, are expressed as: 
\begin{eqnarray}
 |\Psi_{\tau} \rangle &=& \sum_{\kappa} \sum_{\bm{n}'} C_{\kappa \bm{n}'; \tau} |\Phi^{(0)}_{\kappa}\rangle \otimes |\bm{n}'\rangle, 
\label{Eq:Psi}
\end{eqnarray}
where, $C_{\kappa \bm{n}'; \tau}$ are coefficients.
The vibronic state $\tau$ is also characterized by the principal quantum number, irrep and its component. 

\begin{figure}[tb]
\includegraphics[width=7cm]{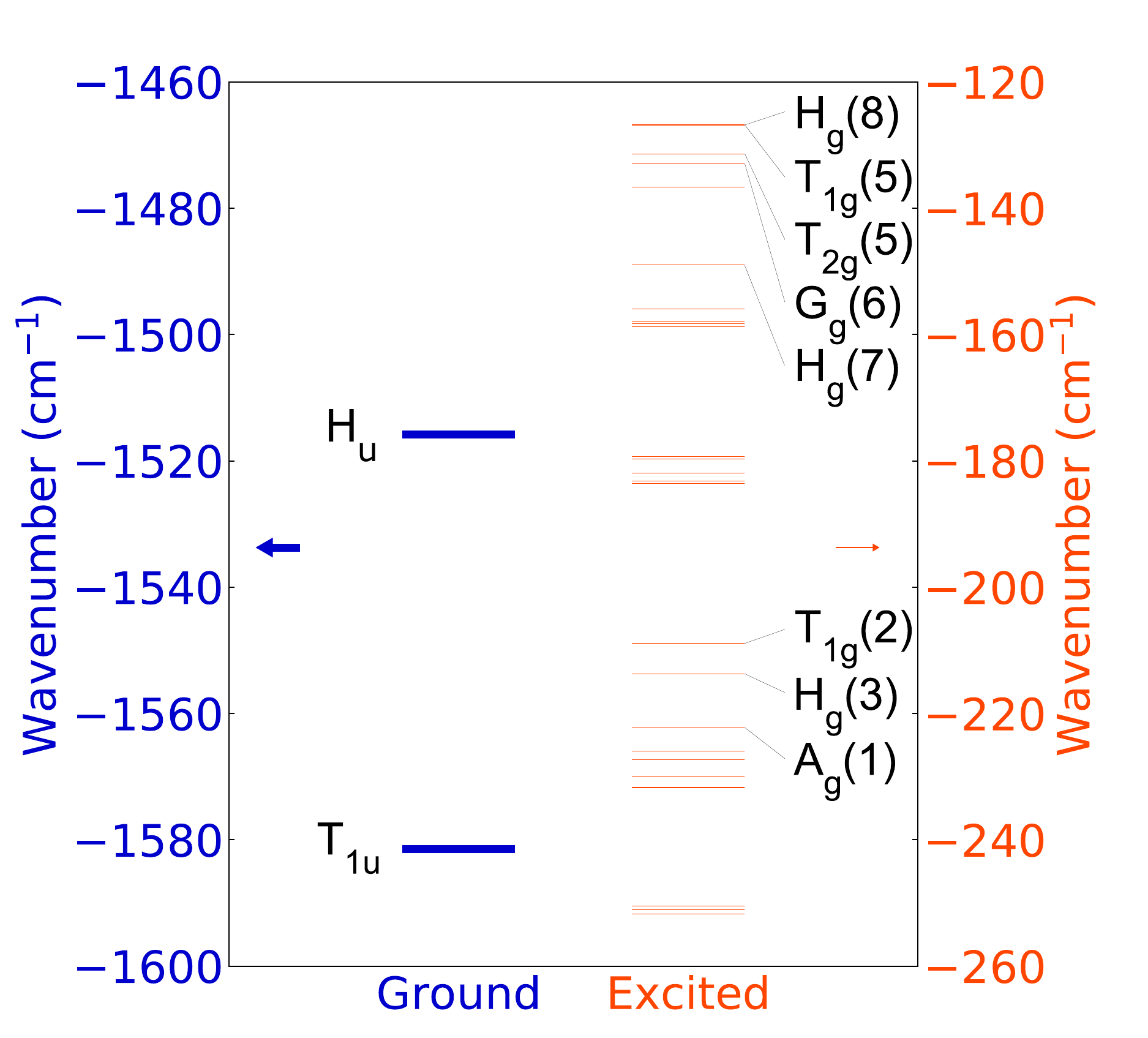}
\caption{
Low-energy vibronic levels mainly originating from the lowest JT states. 
Left column shows the lowest JT levels without excitation of the IR modes. 
The right column shows the vibronic levels with one vibrational excitation
of the IR modes. 
}
\label{Fig:E}
\end{figure}

The vibronic states were numerically derived (Appendix \ref{A:vibronic}). 
The obtained low-energy vibronic levels are shown in Fig. \ref{Fig:E} (see also Fig. S3 \cite{SM}).
The lowest $T_{1u}$ and $H_u$ JT levels are not affected by $\hat{H}_\text{vibro}^{(2)}$, and thus, they are continued to be called JT states (left column in Fig. \ref{Fig:E})
The excited vibronic states arising from the
$T_{1u}$ and $H_u$ JT states (right column in Fig. \ref{Fig:E}) are in a good approximation described by the symmetrized products of $T_{1u}/H_u$ JT states and $|\bm{n}'\rangle$ involving only one IR vibrational excitation (see Appendix \ref{A:Psi} and Table S5 \cite{SM}).

\begin{figure*}[tb]
\begin{tabular}{llll}
(a) & (b) & (c) & (d) \\
\includegraphics[width=4.4cm]{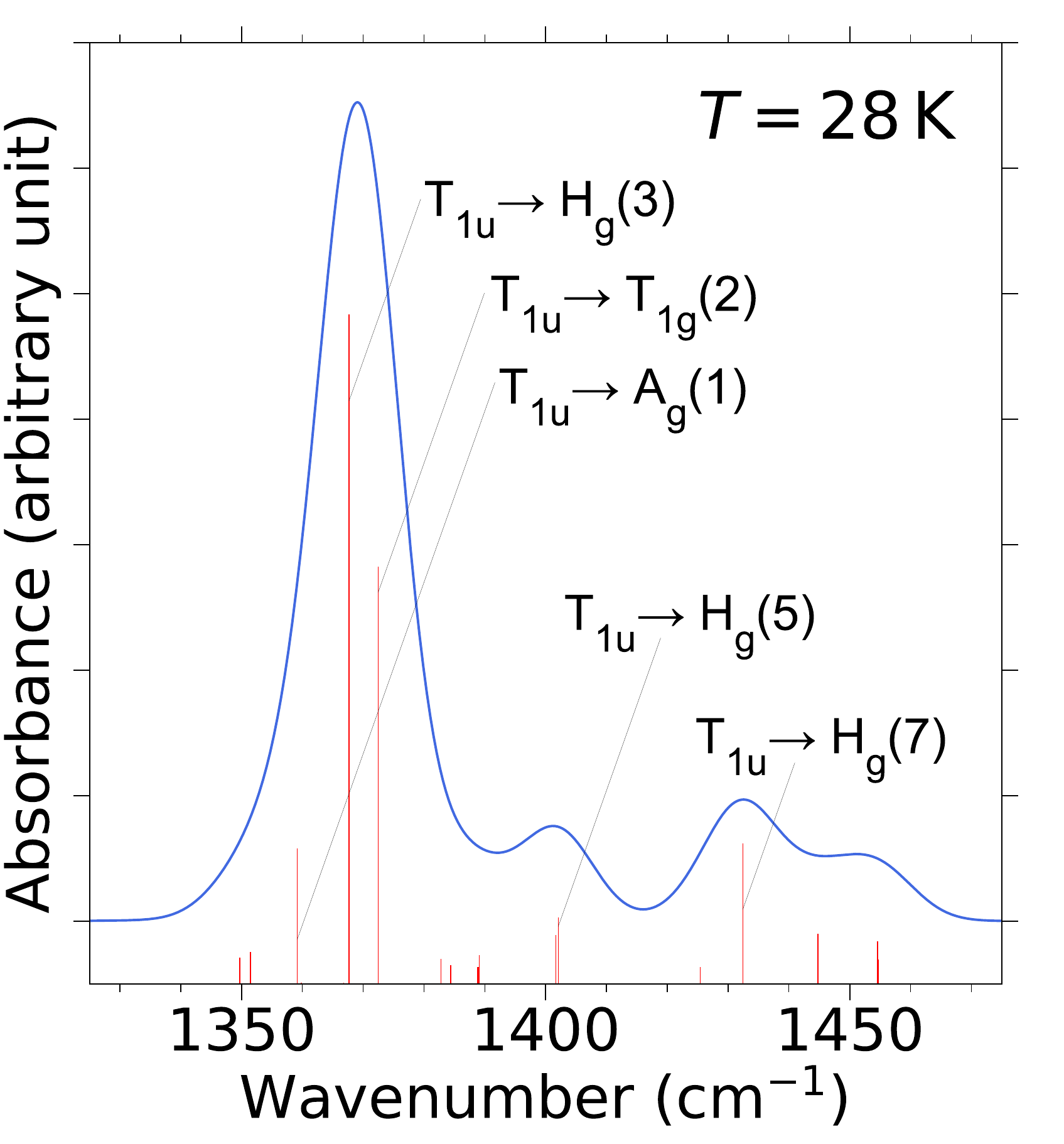}
& 
\includegraphics[width=4.4cm]{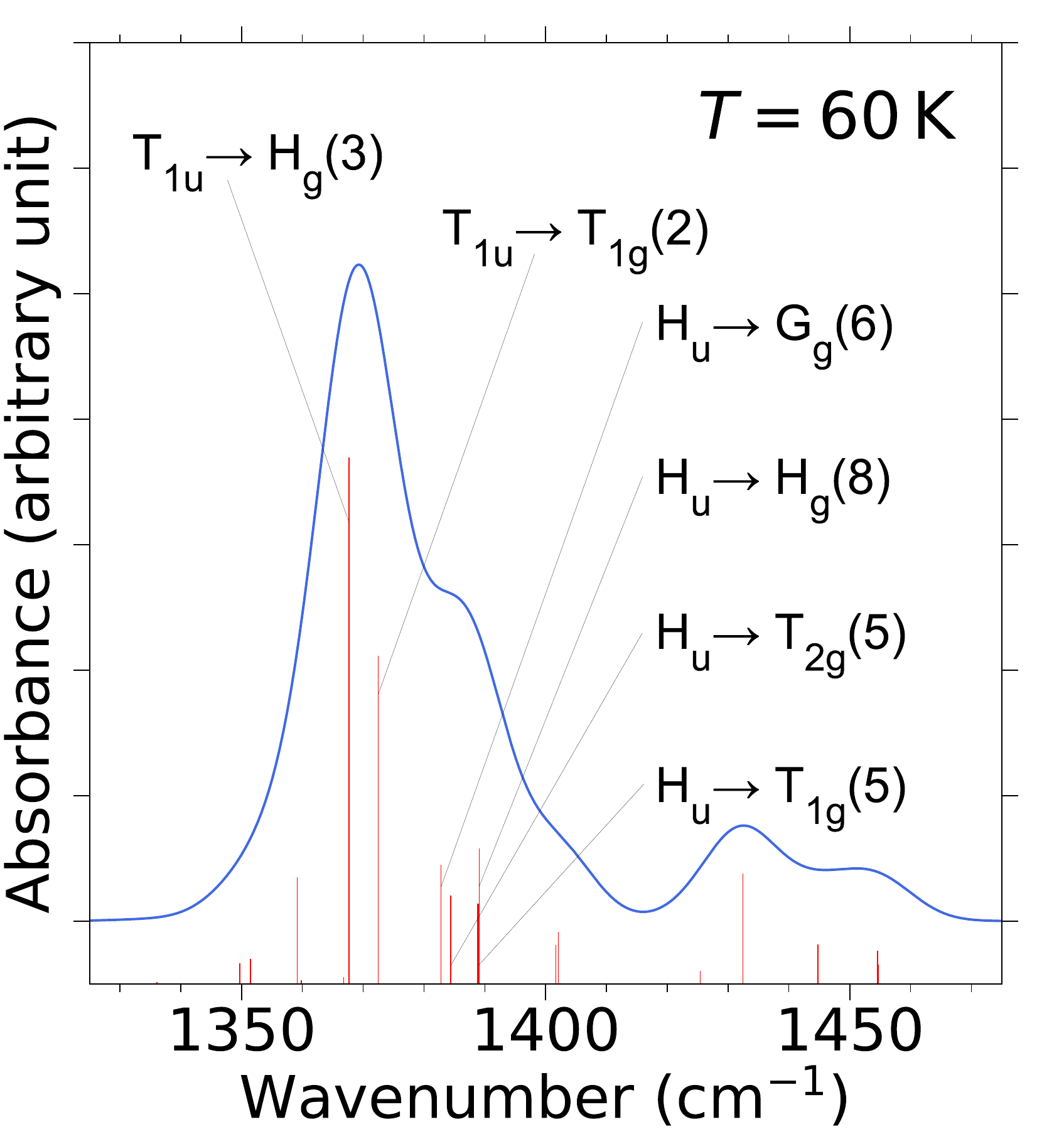}
&
\includegraphics[width=4.4cm]{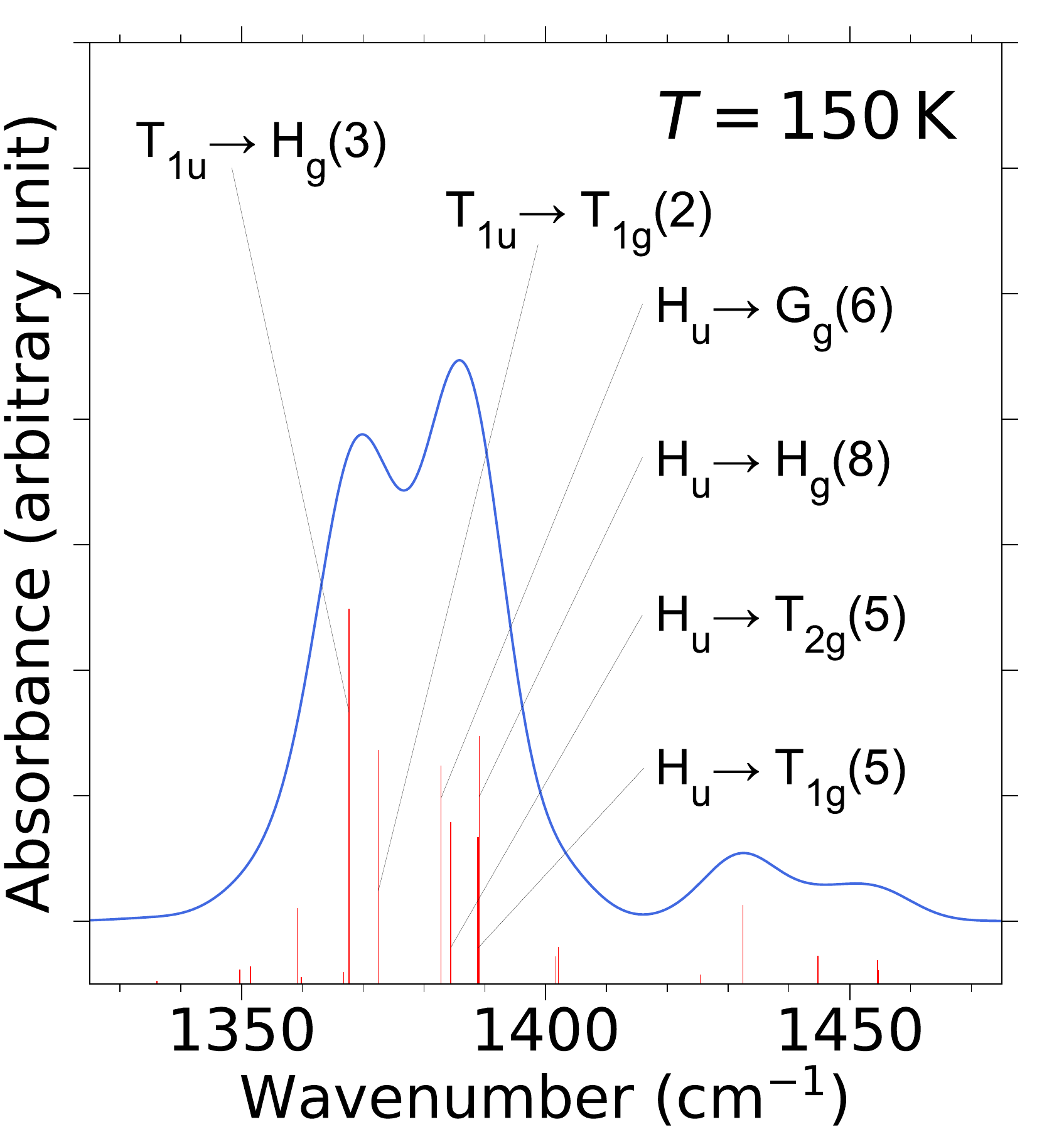}
&
\includegraphics[width=4.2cm]{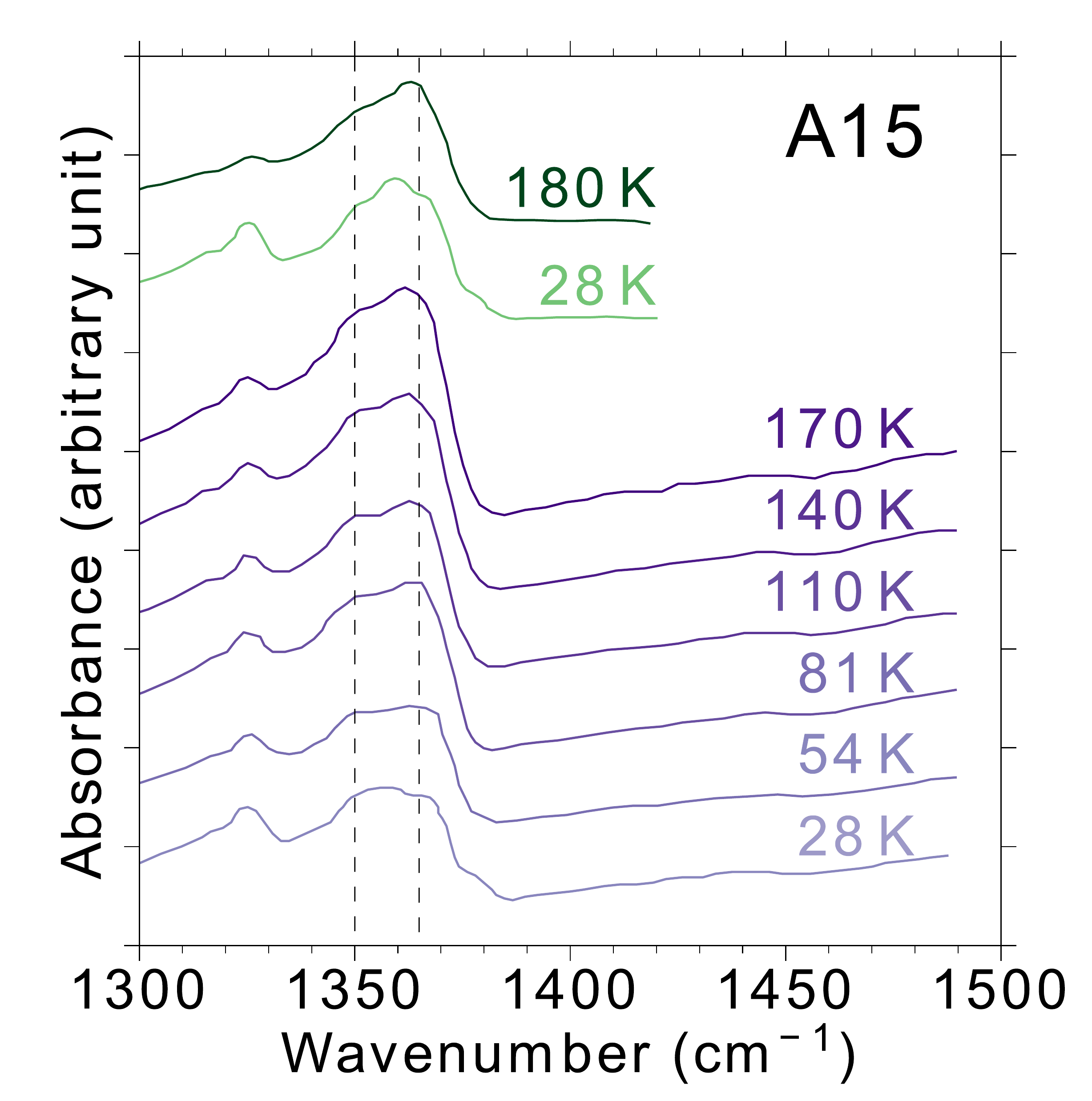}
\end{tabular}
\caption{
(a)-(c) Theoretical and (d) experimental IR spectra of A15 Cs$_3$C$_{60}$. 
Calculated IR spectra of C$_{60}^{3-}$ at (a) 28 K, (b) 60 K, and (c) 150 K. 
All intensities (red vertical lines) are integrated with a Gauss function of full width of half maxima = 14 $\mathrm{cm^{-1}}$ (blue).
In (d), the upper two (green) and lower six (purple) spectra are taken from Ref. \cite{Klupp2012} and Ref.\cite{Kamaras2013}, respectively. 
The vertical dashed lines at 1350 and 1365 cm$^{-1}$ show the positions of the peaks of our interest. 
}
\label{Fig:IR}
\end{figure*}

\section{Vibronic excitations in infrared spectra}
The IR absorption corresponds to the transition from the JT to the excited vibronic states (left and right columns of Fig. \ref{Fig:E}, respectively).
Introducing the coupling to the external electric field $\bm{E}$ \cite{Bersuker1989, Kaplan1995},
\begin{eqnarray}
 \hat{H}_\text{IR} = \sum_\mu \sum_{\gamma = x, y, z} Z_\mu \hat{q}_{t_{1u}(\mu) \gamma} E_\gamma,
\end{eqnarray}
where $Z_\mu$ is effective nuclear charge, the IR absorbance is given by \cite{Sakurai1994}: 
\begin{eqnarray}
 I(\omega, T) &\propto& \sum_{\kappa \tau \gamma} \rho_\kappa(T) \omega \left|\langle \Psi_\tau| \hat{q}_{t_{1u}(4) \gamma} |\Psi_\kappa\rangle\right|^2
 \delta\left( \omega - \omega_{\tau\kappa} \right),
\label{Eq:Intensity}
\end{eqnarray}
where, $\kappa$ indicates the low-lying JT states, $\tau$ denotes the vibronic states with one IR excitation, 
$\rho_\kappa(T)$ is the canonical distribution at temperature $T$, and $\hslash \omega_{\tau\kappa} = E_\tau - E_\kappa$.

The simulated IR spectra of C$_{60}^{3-}$ at (a) 28 K, (b) 60 K, and (c) 150 K are shown in Fig. \ref{Fig:IR}.
At 28 K, all the peaks are mainly composed of the excitations from the lowest $T_{1u}$ JT level. 
The highest peak around 1370 cm$^{-1}$ includes three major components corresponding to the excitations from the $T_{1u}$ to $H_{g}(3)$, $T_{1g}(2)$, and $A_{g}(1)$ vibronic levels (Fig. \ref{Fig:E} and Table S6 \cite{SM}). 
In these IR-admixed vibronic states, the products of the
$T_{1u}$ JT state and one $t_{1u}(4)$ vibrational excitation have large weight (Table S5 \cite{SM}). 
As temperature rises, the first excited $H_u$ JT state is also thermally populated, and thus, the fraction of the transitions from the $T_{1u}$ ($H_u$) JT state decreases (increases).
The new peaks observed at high temperature correspond to the transitions from the $H_u$ to the $G_g(6)$, $H_g(8)$, $T_{2g}(5)$ and $T_{1g}(5)$ vibronic levels (Fig. \ref{Fig:E} and Table S6 \cite{SM}
) which mainly originate from the products of the
$H_u$ JT state and one $t_{1u}(4)$ excitation (Table S5 \cite{SM}). 
The contributions from the $H_u$ are not negligible already at 60 K and eventually become stronger at high temperature
[Fig. \ref{Fig:IR} (b), (c)]
\footnote{
At higher temperature above 150 K, many peaks from these excited JT levels above the $T_{1u}$ and $H_u$ ones and the broadening of the line width smear the shape of the fine structures of the IR spectrum. 
}.
The evolution explains the basic features of the experimental data:
in the spectra, the peak around 1350 cm$^{-1}$ from the $T_{1u}$ JT level is the highest at $T = 28$ K, while the peak at 1365 cm$^{-1}$ from the $H_{u}$ JT level becomes higher as temperature increases [Fig. \ref{Fig:IR} (d)]
\footnote{
The slight difference between the theoretical and experimental shapes of the main peaks might be explained by the small shift of the theoretical vibronic levels from the real ones due to e.g. the absence of weak intersite interactions and crystal field effect in our calculations or the imperfect agreement between the DFT and experimental parameters.
One also should note the sample dependence of IR spectra: 
Although both experimental data show the growth of new peak as the rise of temperature, the main peaks at 28 K at the bottom and the top of Fig. \ref{Fig:IR} (d) have slightly different fine structures.
}.
Thus, the present theory proves that the temperature dependence indeed originates from thermal population of vibronic excitation from the $T_{1u}$ to the $H_u$ JT states.
Besides the main peaks, some of the other fine structures can be explained. 
For example, the gradual decrease of the peak at 1400 cm$^{-1}$ [$T_{1u} \rightarrow H_g(5)$] with the increase of temperature resembles the temperature dependence of the shoulder at 1380 cm$^{-1}$ in the experimental data [Fig. \ref{Fig:IR} (d)].
Note that such small structures arises due to the quadratic vibronic coupling involving 
IR inactive $g_u/h_u$ modes besides IR active $t_{1u}$.

\section{Discussion}
\label{Sec:Discussion}
The interpretation of the IR spectra of JT systems is often based on the modification of selection rules due to the symmetry lowering \cite{Thorson1958}.
Thus, in the previous quantum chemistry study of Cs$_3$C$_{60}$ cluster, the IR spectra were calculated for some electronic configurations at statically deformed structure \cite{Naghavi2016}.
Considering the complicated orbital-lattice entangled nature of vibronic states, such simplified treatment is not sufficient to reveal the mechanisms of the implication of dynamical JT effect in the observed properties. 
The theoretical framework for a complete treatment of the JT dynamics and IR spectra in high symmetric JT systems is established in this work. 
On this basis, it is shown that the assumption of a change of adiabatic potential energy surface
by thermal expansion \cite{Klupp2012, Kamaras2013} is not necessary for the description of the temperature evolution of IR spectra. 

The obtained results can be applied to the analysis of the IR spectra in fullerides close to Mott transition where the IR spectra are practically unaffected by weak intersite interactions. 
In the metallic phase, the electron transfer interaction $\hat{H}_\text{t}$ is not negligible.
In the ground state, $\hat{H}_\text{t}$ forms continuous vibronic states without quenching the JT dynamics because of the large stabilization energy of the latter \cite{Iwahara2015}.
As a consequence, the spin gap induced by the JT dynamics is observed even in the fullerides with small lattice constant \cite{Brouet2002c}.
The IR-admixed vibronic states (right column of Fig. \ref{Fig:E}) are also mixed with many excited vibronic states involving one IR vibrational excitation
by $\hat{H}_\text{t}$.
At a sufficiently strong interaction of the excited vibronic states with a continuum of bulk states, a Fano resonance in IR spectra will appear. 
We stress that the modification of the shape of IR spectra into a Fano resonance \cite{Zadik2015}
does not necessarily mean the quench of JT effect in the ground state. 

The present work can also contribute to the description of the light-induced superconductivity in K$_3$C$_{60}$ \cite{Mitrano2016, Cantaluppi2017}.
The model used in recent works \cite{Kennes2017, Sentef2017} for the discussion of possible mechanism based on a simplified quadratic vibronic coupling, should be replaced with Eq. (\ref{Eq:H}). 
In Ref.  \cite{Mitrano2016}, the importance of the cubic coupling was also proposed to relate the vibration of IR active $t_{1u}$
mode and JT distortion, while it was shown here that these lattice degrees of freedom are already entangled (\ref{Eq:Psi}) within the quadratic vibronic model via degenerate electronic terms.

\section{Conclusion}
The quantum mechanical framework for the description of the IR spectra of dynamical JT system is developed. 
Based on the first-principles calculations, the nonlinear vibronic Hamiltonian of C$_{60}^{3-}$ site was derived, and the vibronic states involving both the JT and IR degrees of freedom were calculated.
On this basis, the relation between the temperature evolution of IR spectra in Mott insulating Cs$_3$C$_{60}$ and JT dynamics was established. 
The first-principles calculations also showed non-negligible nonlinear vibronic couplings between the IR active $t_{1u}$ mode and non-IR active $g_u$ and $h_u$ modes, 
giving rise to the fine structures of the IR spectra. 
It is worthy to note that, contrary to the previous approaches, static JT distortions are not assumed to be responsible for the temperature evolution of IR spectra as well as their fine structure. 
Because the fine structure of the IR spectra is governed by the IR-admixed vibronic structure, the reconsideration on the relation between the IR spectra in metallic fullerides and the nature of JT dynamics as function of their volume is required. 
The developed theoretical approach can be applied to the study of the spectroscopic and light-induced electronic properties of various correlated dynamical JT materials.

\section*{Acknowledgment}
We would like to thank Henri Alloul, Katalin Kamar\'{a}s and Erio Tosatti for useful discussions. 
N.I. is supported by Japan Society for the Promotion of Science Overseas Research Fellowship.
K.T. acknowledges the Scientific Research Funds, the World Premier International Research Center Initiative (WPI) from Ministry of Education, Culture, Sports, Science and Technology (MEXT) of Japan and the Grant in Aid for Scientific Research (No. 17-18H05326, 18H04304, 18H03883, 18H03858) and thermal management of CREST, JST.

\appendix
\section{Quadratic vibronic coupling}
In this section, SO(3) symmetry and the spherical basis for the $t_{1u}$ and $h_{u/g}$ irreps. of $I_h$ group are used. 
The Condon-Shortley's phase convention is taken for the spherical harmonics \cite{Varshalovich1988}.

\subsection{Derivation of Eq. (\ref{Eq:HQJT})}
\label{A:derivation}
The $t_{1u}$ LUMOs couple to the symmetrized $h_g$ polynomials of the normal coordinates.
The quadratic $h_g$ polynomials are calculated as 
\begin{eqnarray}
 \{ \hat{q}_{\Gamma_1(\mu_1)} \otimes \hat{q}_{\Gamma_2(\mu_2)} \}_{\nu h_g \gamma} 
 &=& 
 \sum_{\gamma_1 \gamma_2} 
 \langle \nu h_g \gamma| \Gamma_1 \gamma_1 \Gamma_2 \gamma_2\rangle 
\nonumber\\
 &\times&
 \hat{q}_{\Gamma_1(\mu_1)\gamma_1} \hat{q}_{\Gamma_2(\mu_2)\gamma_2},
\label{Eq:qq}
\end{eqnarray}
where the sum is over the components of the irreps. $\Gamma_1$ and $\Gamma_2$. 
See for the explicit form Sec. II in Ref. \cite{SM}. 
The quadratic vibronic coupling is expressed as 
\begin{eqnarray}
 \hat{H}_\text{vibro}^{(2)} &=& \sum_{\mu_i \Gamma_i} \sum_{\nu \gamma} \sum_{\lambda \lambda' \sigma}
 V_{\nu \mu_1 \mu_2}^{\Gamma_1 \Gamma_2} 
 \{ \hat{q}_{\Gamma_1(\mu_1)} \otimes \hat{q}_{\Gamma_2(\mu_2)} \}_{\nu h_g \gamma} 
\nonumber\\
 &\times&
 \sqrt{\frac{5}{2}} \langle t_{1u} \lambda | t_{1u} \lambda', h_g \gamma \rangle 
 \hat{c}_{t_{1u}\lambda \sigma}^\dagger \hat{c}_{t_{1u}\lambda' \sigma},
\label{Eq:HQJT_2nd}
\end{eqnarray}
where $\sigma$ indicates the projection of electron spin and $\hat{c}_{t_{1u}\lambda \sigma}^\dagger$ and $\hat{c}_{t_{1u}\lambda \sigma}$ are the electron creation and annihilation operators in spin-orbital $t_{1u} \lambda \sigma$.
Transforming the real components of the irreps. into the spherical ones, Eq. (\ref{Eq:HQJT_2nd}) reduces to the same form as Eq. (A1) in Ref. \cite{Liu2018}. 
Thus, by the same procedure as described in the Appendix A in Ref. \cite{Liu2018}, we obtain the spherical form of Eq. (\ref{Eq:HQJT}):
\begin{eqnarray}
 \hat{H}^{(2)}_\text{vibro}
 &=& 
  \sum_{\mu_i \Gamma_i} \sum_{\nu m''} 
  \sum_{nn'} (-1)^{m''}
  V^{\Gamma_1 \Gamma_2}_{\nu \mu_1 \mu_2}
\nonumber\\
 &\times& 
 \{\hat{q}_{\Gamma_1(\mu_1)} \otimes \hat{q}_{\Gamma_2(\mu_2)} \}_{\nu 2,-m''}
 \sqrt{\frac{15}{2}}
 \langle 1n|2n', 2m'' \rangle
\nonumber\\
 &\times& 
 \left[
  |1n \rangle \langle 2n'| - (-1)^{m''} |2,-n'\rangle \langle 1, -n|
 \right],
\label{Eq:HQJT_sph}
\end{eqnarray}
where, $|1n\rangle$ and $|2n'\rangle$ express the spherical form of the $T_{1u}$ and $H_u$ multiplet states, respectively, and $n,n',m''$ the projections of the angular momenta. 
By the inverse transformation from the spherical into the Cartesian basis, we obtain Eq. (\ref{Eq:HQJT}). 
The difference in the coefficients in Eqs. (\ref{Eq:HQJT}), (\ref{Eq:HQJT_sph}) and Eq. (\ref{Eq:HQJT_2nd}) by $\sqrt{3}$ comes from the reduced matrix element of Racah's $\hat{U}^{(2)}$ operator. 
Since the spin and orbital degrees of freedom are independent from each other, Eqs. (\ref{Eq:HQJT}) and (\ref{Eq:HQJT_sph}) show only the orbital part.

\subsection{Model Hamiltonian for the IR modes}
\label{A:HQJT}
One-electron quadratic vibronic Hamiltonian for the IR modes is given by 
\begin{eqnarray}
 \hat{H}_0' + \hat{H}_\text{vibro}^{(2)\prime} &=& 
 \left(|t_{1u}x\rangle, |t_{1u}y\rangle, |t_{1u}z\rangle  \right)
\nonumber\\
&\times&
 \left(\mathbf{H}_{0}' + \mathbf{H}^{(2)\prime}_\text{vibro}\right)
 \begin{pmatrix}
  \langle t_{1u}x|\\ 
  \langle t_{1u}y|\\ 
  \langle t_{1u}z|  
 \end{pmatrix}.
\label{Eq:HnonJT}
\end{eqnarray}
Here, the projection of the electron spin is omitted, $\mathbf{H}_{0}'$ involves the orbital energy $\epsilon_{t_{1u}}$ and the harmonic oscillator terms, and $\mathbf{H}^{(2)\prime}_\text{vibro}$ is the quadratic vibronic coupling:
\begin{widetext}
\begin{eqnarray}
 \mathbf{H}_{0}' &=&
 \left(
 \epsilon_{t_{1u}} 
+  
  \sum_{\gamma = x, y, z} \frac{k_{t_{1u}(4)}}{2} q_{t_{1u}(4) \gamma}^2 
+ 
  \sum_{\gamma = a, x, y, z} \frac{k_{g_u(5)}}{2} q_{g_u(5) \gamma}^2 
+ 
  \sum_{\gamma = \theta, \epsilon, \xi, \eta, \zeta} \frac{k_{h_u(6)}}{2} q_{h_u(6) \gamma}^2 
 \right) 
\mathbf{I},
\label{Eq:H0_mat}
\\
 \mathbf{H}^{(2)\prime}_\text{vibro} 
 &=&
 \sum_{\gamma = \theta, \epsilon, \xi, \eta, \zeta} 
 \left(
   V^{t_{1u} t_{1u}}_{44} \{\hat{q}_{t_{1u}(4)} \otimes \hat{q}_{t_{1u}(4)} \}_{h_g \gamma} 
 + V^{g_u g_u}_{55} \{\hat{q}_{g_u(5)} \otimes \hat{q}_{g_u(5)} \}_{h_g \gamma} 
 + \sum_{\nu=1,2} V^{h_u h_u}_{\nu 66} \{\hat{q}_{h_u(6)} \otimes \hat{q}_{h_u(6)}\}_{\nu h_g \gamma} 
 \right.
\nonumber\\
 &+&
 \left.
   2V^{t_{1u} g_u}_{45} \{\hat{q}_{t_{1u}(4)} \otimes \hat{q}_{g_u(5)}\}_{h_g\gamma} 
 + 2V^{t_{1u} h_u}_{46} \{\hat{q}_{t_{1u}(4)} \otimes \hat{q}_{h_u(6)}\}_{h_g\gamma} 
 +
  \sum_{\nu=1,2} 2V^{g_u h_u}_{\nu 56} \{\hat{q}_{g_u(5)} \otimes \hat{q}_{h_u(6)}\}_{\nu h_g\gamma} 
 \right)
 \mathbf{M}_\gamma,
\label{Eq:H2_mat}
\end{eqnarray}
\end{widetext}
$k_{\Gamma(\mu)}$ indicates the elastic parameter, $\mathbf{I}$ is the three-dimensional unit matrix, and 
\begin{eqnarray}
 \mathbf{M}_\theta &=& 
 \begin{pmatrix}
   \frac{1}{2} & 0 & 0 \\
   0 & \frac{1}{2} & 0 \\
   0 & 0 & -1 \\
 \end{pmatrix},
\quad 
 \mathbf{M}_\epsilon = 
 \begin{pmatrix}
   -\frac{\sqrt{3}}{2} & 0 & 0 \\
   0 & \frac{\sqrt{3}}{2} & 0 \\
   0 & 0 & 0 \\
 \end{pmatrix},
\nonumber\\
 \mathbf{M}_\xi &=& 
 \begin{pmatrix}
   0 & 0 & 0 \\
   0 & 0 & -\frac{\sqrt{3}}{2} \\
   0 & -\frac{\sqrt{3}}{2} & 0 \\
 \end{pmatrix},
\quad 
 \mathbf{M}_\eta = 
 \begin{pmatrix}
   0 & 0 & -\frac{\sqrt{3}}{2} \\
   0 & 0 & 0 \\
   -\frac{\sqrt{3}}{2} & 0 & 0 \\
 \end{pmatrix},
\nonumber\\
 \mathbf{M}_\zeta &=& 
 \begin{pmatrix}
   0 & -\frac{\sqrt{3}}{2} & 0 \\
   -\frac{\sqrt{3}}{2} & 0 & 0 \\
   0 & 0 & 0 \\
 \end{pmatrix}.
\end{eqnarray}
In Eqs. (\ref{Eq:H0_mat}) and (\ref{Eq:H2_mat}), only the IR modes \cite{Note5} are included. 

\subsection{Derivation of Eq. (\ref{Eq:Hquad})}
\label{A:MatHQJT}
The transformation from Eq. (\ref{Eq:HQJT}) to Eq. (\ref{Eq:Hquad}) is described here. 
In this subsection, to make use of the SO(3) symmetry of the linear $t_{1u}^3 \otimes h_g$ JT system, the vibronic angular momentum $J$ is used instead of real irrep $\Gamma$ of $I_h$ group. 

Using the identity operator for the JT states, $\hat{I}^{(0)} = \sum_{\alpha J m P} |\Phi^{(0)}_{\alpha J m P}\rangle \langle \Phi^{(0)}_{\alpha J m P}|$,
\begin{eqnarray}
 \hat{H}_\text{vibro}^{(2)} &=& 
 \hat{I}^{(0)} \hat{H}_\text{vibro}^{(2)} \hat{I}^{(0)} 
\nonumber\\
 &=&
 \sum_{\alpha J m P} 
 \sum_{\alpha' J' m' P'} 
  \langle \Psi^{(0)}_{\alpha J m P}|\hat{H}_\text{vibro}^{(2)}|\Psi^{(0)}_{\alpha' J' m' P'}\rangle 
\nonumber\\
 &\times&
 |\Psi^{(0)}_{\alpha J m P}\rangle \langle \Psi^{(0)}_{\alpha' J' m' P'}|,
\label{Eq:IHQI}
\end{eqnarray}
where, $J$ is the vibronic angular momentum and $m$ is its projection. 
Substituting Eq. (\ref{Eq:HQJT_sph}) into the matrix element of Eq. (\ref{Eq:IHQI}), we obtain  
\begin{eqnarray}
 \hat{H}_\text{vibro}^{(2)}
 &=& 
 \sum_{\alpha J m P} 
 \sum_{\alpha' J' m' P'} 
 \sum_{\mu_i \Gamma_i} \sum_{\nu m''} 
 V^{\Gamma_1 \Gamma_2}_{\nu \mu_1 \mu_2}
\nonumber\\
 &\times&
 (-1)^{m''} \{\hat{q}_{\Gamma_1(\mu_1)} \otimes \hat{q}_{\Gamma_2(\mu_2)} \}_{\nu 2,-m''}
 \sqrt{\frac{15}{2}}
\nonumber\\
 &\times& 
 \langle \Psi^{(0)}_{\alpha J m P}|
 \hat{M}_{2m''}
  |\Psi^{(0)}_{\alpha' J' m' P'}\rangle
 |\Psi^{(0)}_{\alpha J m P}\rangle \langle \Psi^{(0)}_{\alpha' J' m' P'}|,
\nonumber\\
\label{Eq:IHQI2}
\end{eqnarray}
with rank 2 irreducible tensor operator 
\begin{eqnarray}
 \hat{M}_{2m''} &=&
  \sum_{n n'} 
 \langle 1 n|2 n', 2 m'' \rangle
\nonumber\\
 &\times&
 \left[ |1 n \rangle \langle 2 n'| - (-1)^{m''} |2, -n' \rangle \langle 1, -n| \right].
\end{eqnarray}
Applying Wigner-Eckart theorem \cite{Varshalovich1988} to the matrix elements of $\hat{M}_{2 m''}$, 
\begin{eqnarray}
 \langle \Psi^{(0)}_{\alpha J m +}|
&&
 \hat{M}_{2 m''} |\Psi^{(0)}_{\alpha' J' m' -}\rangle
\nonumber\\
&&=
 \frac{
 \langle \Psi^{(0)}_{\alpha J +}\Vert \hat{M}_{2} \Vert \Psi^{(0)}_{\alpha' J' -}\rangle}
 {\sqrt{2J+1}} 
 \langle J m | J' m', 2 m'' \rangle
\nonumber\\
 &&=
 \Xi(\alpha J +; \alpha' J' -) 
 \langle J m | J' m', 2 m'' \rangle.
\label{Eq:Xi}
\end{eqnarray}
Here, the parities of the $T_{1u}$ and $H_u$ terms are defined to be $+1$ and $-1$, respectively [see Eq. (11) in Ref. \cite{Liu2018}], and $\Xi(\alpha J P; \alpha' J' P')$ is defined by 
\begin{eqnarray}
 \Xi(\alpha J P; \alpha' J' P') 
 &=&
 \frac{
 \langle \Psi^{(0)}_{\alpha J P}\Vert \hat{M}_{2} \Vert \Psi^{(0)}_{\alpha' J' P'}\rangle}
 {\sqrt{2J+1}}.
\end{eqnarray}
Therefore, by the substitution of Eq. (\ref{Eq:Xi}) into Eq. (\ref{Eq:IHQI2}), we obtain 
\begin{eqnarray}
 \hat{H}_\text{vibro}^{(2)}
 &=& 
 \sum_{\alpha J m } 
 \sum_{\alpha' J' m' } 
  \sum_{\mu_i \Gamma_i} \sum_{\nu m''} 
 V^{\Gamma_1 \Gamma_2}_{\nu \mu_1 \mu_2}
 \Xi(\alpha J +; \alpha' J' -) 
\nonumber\\
 &\times&
 (-1)^{m''} \{\hat{q}_{\Gamma_1(\mu_1)} \otimes \hat{q}_{\Gamma_2(\mu_2)} \}_{\nu 2,-m''}
 \sqrt{\frac{15}{2}}
\nonumber\\
 &\times&
 \langle J m | J' m', 2 m'' \rangle
 |\Psi^{(0)}_{\alpha J m +}\rangle \langle \Psi^{(0)}_{\alpha' J' m' -}| + \text{h.c.}
\nonumber\\
 &=& 
 \sum_{\alpha J m P} 
 \sum_{\alpha' J' m' P'} 
  \sum_{\mu_i \Gamma_i} \sum_{\nu m''} 
 V^{\Gamma_1 \Gamma_2}_{\nu \mu_1 \mu_2}
 \Xi(\alpha J P; \alpha' J' P') 
\nonumber\\
 &\times&
 (-1)^{m''} \{\hat{q}_{\Gamma_1(\mu_1)} \otimes \hat{q}_{\Gamma_2(\mu_2)} \}_{\nu 2,-m''}
 \sqrt{\frac{15}{2}}
\nonumber\\
 &\times&
 \langle J m | J' m', 2 m'' \rangle
 |\Psi^{(0)}_{\alpha J m P}\rangle \langle \Psi^{(0)}_{\alpha' J' m' P'}|.
\label{Eq:Hquad_J}
\end{eqnarray}
To obtain the second form, 
\begin{eqnarray}
\left[\{\hat{q}_{\Gamma_1(\mu_1)}  \right. && \otimes \left. \hat{q}_{\Gamma_2(\mu_2)} \}_{\nu 2,-m''} \right]^\dagger 
\nonumber\\
 &&= 
(-1)^{m''} \{\hat{q}_{\Gamma_1(\mu_1)} \otimes \hat{q}_{\Gamma_2(\mu_2)} \}_{\nu 2m''},
\end{eqnarray}
and 
\begin{eqnarray}
 \Xi(&&\alpha J +; \alpha' J' -) 
 \langle J m | J' m', 2 m'' \rangle
\nonumber\\
 &&=  
 \Xi(\alpha' J' -;\alpha J +) 
 (-1)^{m''} \langle J'm' | Jm, 2,-m'' \rangle,
\end{eqnarray}
are used. 
Nonzero $\Xi(\alpha J +; \alpha' J' -)$'s for the low-energy JT levels are shown in Table S4 \cite{SM}.
Finally, using the relation between the irreps. of SO(3) and $I_h$ in Eq. (\ref{Eq:Hquad_J}), we obtain Eq. (\ref{Eq:Hquad}). 
For details of the relation of these two representations, see Sec. I A in the Supplemental Materials of Ref. \cite{Liu2018Q}.

\section{Computational methods}
\subsection{DFT calculations}
\label{A:DFT}
In order to determine the frequencies and the quadratic vibronic coupling parameters, DFT calculations of Cs$_3$C$_{60}$ cluster were performed. 
The cluster con sits of one C$_{60}$, the nearest 12 Cs atoms and several thousands of point charges for distant C$_{60}^{3-}$ ($-3$) and Cs$^+$ ($+1$).  
For the calculations, B3LYP hybrid exchange-correlation functional was used with a triple-zeta polarization 6-311G(d) and double zeta 3-21G basis sets for C and Cs, respectively \cite{g09long}.
In order to avoid the artificial splitting of the $t_{1u}$ orbital levels, the $^4A_u$ electronic configuration was treated within spin-unrestricted approach. 
The quadratic vibronic coupling constants were derived by fitting the $t_{1u}$ LUMO levels with respect to the deformed structures to the model Hamiltonian, Eq. (\ref{Eq:HnonJT}). 
The results of the fitting are shown in Fig. S2 \cite{SM}. 

\subsection{Calculation of vibronic states}
\label{A:vibronic}
The vibronic Hamiltonian matrix was derived using the JT states of Ref. \cite{Liu2018}, and then numerically diagonalized
\footnote{The program for the calculation of \unexpanded{$|\Psi_\tau\rangle$} and IR spectra is available at https://github.com/quasi-human.}.
The basis of $|\Psi_\tau\rangle$ (\ref{Eq:Psi}) was truncated on the basis of the excitation energy of JT states and the number of vibrational excitations for IR modes \cite{Note5}: the former was $<$ 90 meV and the latter was 3,
\begin{eqnarray}
0 \le  \sum_{\Gamma(\mu) \gamma} n'_{\Gamma(\mu)\gamma}  \le 3,
\end{eqnarray}
where, the sum is over the IR modes, and $n'_{\Gamma(\mu)\gamma} \ge 0$ is vibrational quantum number.
The cutoff of the JT states restricts the maximum temperature for the simulation of the IR spectra as in the case of spin gap \cite{Liu2018} because the JT levels higher than the cutoff cannot be populated.
The maximum temperature treated in this work is 150 K.

\section{Assignment of the vibronic states} 
\label{A:Psi}
The lowest vibronic states (left column of Fig. \ref{Fig:E}) are expressed by the linear combinations of the products of the JT states and the vibrational vacuum $|\bm{n}' = \bm{0}\rangle$ or two vibrationally excited states of the IR modes \cite{Note5}. 
The contribution from the latter is negligible in the present case, and thus, these low-energy vibronic states are well defined by the JT states. 

On the other hand, the excited vibronic states (right column of Fig. \ref{Fig:E}) are described by the linear combination of the products of the JT states and the one- or three-vibrationally excited IR states, and the main contributions come from the former.  
Therefore, the excited vibronic states $|\Psi_\tau\rangle$ are assigned by the symmetrized products of the JT and the one vibrationally excited states.
The symmetrized states are expressed as
\begin{eqnarray}
 |(\Gamma_1 \Gamma_2) n\Gamma \gamma\rangle &=& \sum_{\gamma_1 \gamma_2} |\Gamma_1 \gamma_1\rangle \otimes |\Gamma_2 \gamma_2\rangle
 \langle \Gamma_1 \gamma_1, \Gamma_2 \gamma_2|\nu\Gamma \gamma\rangle,
\nonumber\\
\label{Eq:symm}
\end{eqnarray}
where, $\Gamma_1$ and $\Gamma_2$ indicate the irreps. of the JT and IR states, respectively. 
Thus, $\Gamma_2$ is the irrep of the excited IR mode. 
The product states of the $T_{1u}$ JT state and the $t_{1u}$, $g_u$, $h_u$ vibrational states split into 10 states,
\begin{eqnarray}
 T_{1u} \otimes t_{1u} &=& A_g \oplus T_{1g} \oplus H_g, 
\nonumber\\
 T_{1u} \otimes g_u &=& T_{2g} \oplus G_g \oplus H_g, 
\nonumber\\
 T_{1u} \otimes h_u &=& T_{1g} \oplus T_{2g} \oplus G_g \oplus H_g,
\end{eqnarray}
and those with the $H_u$ JT state into 16 states,
\begin{eqnarray}
 H_u \otimes t_{1u} &=& T_{1g} \oplus T_{2g} \oplus G_g \oplus H_g, 
\nonumber\\
 H_u \otimes g_u &=& T_{1g} \oplus T_{2g} \oplus G_g \oplus 2H_g, 
\nonumber\\
 H_u \otimes h_u &=& A_g \oplus T_{1g} \oplus T_{2g} \oplus 2G_g \oplus 2H_g,
\end{eqnarray}
respectively. 
The explicit forms of the symmetrized vibronic states (\ref{Eq:symm}) are given in Sec. IV A of Ref. \cite{SM}. 
The contribution of each symmetrized state (\ref{Eq:symm}) to the vibronic states $|\Psi_{\tau} \rangle$, Eq. (\ref{Eq:Psi}), is quantitatively evaluated by calculating 
\begin{eqnarray}
 w(\Gamma_1 \Gamma_2 \nu \Gamma, \tau) &=& 
 \langle \Psi_{\tau}| \hat{I}_{(\Gamma_1\Gamma_2)\nu\Gamma} |\Psi_{\tau} \rangle,
\label{Eq:w}
\end{eqnarray}
where, $\hat{I}_{(\Gamma_1\Gamma_2)\nu\Gamma}$ is the projector into the symmetrized states $(\Gamma_1 \Gamma_2) \nu \Gamma \gamma$, 
\begin{eqnarray}
 \hat{I}_{(\Gamma_1\Gamma_2)\nu\Gamma} &=& \sum_{\gamma} |(\Gamma_1 \Gamma_2) \nu\Gamma \gamma\rangle \langle (\Gamma_1 \Gamma_2) \nu\Gamma \gamma|.
\end{eqnarray}
Eq. (\ref{Eq:w}) also enables us to skip the analysis of the structure of $|\Psi_\tau\rangle$ for the assignment.
The list of $w(\Gamma_1 \Gamma_2 \nu \Gamma, \tau)$ are given in Table S5 \cite{SM}.


%

\end{document}